\documentclass[twocolumn,showpacs,superscriptaddress,email,floatfix,longbibliography,prb]{revtex4-2}

\usepackage{amsmath}
\usepackage{inputenc}
\usepackage{graphicx}
\usepackage{braket}
\usepackage{longtable}
\usepackage{bbold}
\usepackage{amsmath,amssymb,amsfonts,bbm,graphicx,hyperref,color}
\usepackage{multirow}
\usepackage{graphicx}
\usepackage[table,xcdraw]{xcolor}
\usepackage{natbib}
\usepackage{hyperref}
\usepackage{xr}

\begin{document}
\title{Signatures of a quantum stabilized fluctuating phase and critical dynamics in a kinetically-constrained open  many-body system with two absorbing states}
\author{Federico Carollo}
\affiliation{Institut f\"ur Theoretische Physik, Universit\"at T\"ubingen, Auf der Morgenstelle 14, 72076 T\"ubingen, Germany}
\author{Markus Gnann}
\affiliation{Institut f\"ur Theoretische Physik, Universit\"at T\"ubingen, Auf der Morgenstelle 14, 72076 T\"ubingen, Germany}
\author{Gabriele Perfetto}
\affiliation{Institut f\"ur Theoretische Physik, Universit\"at T\"ubingen, Auf der Morgenstelle 14, 72076 T\"ubingen, Germany}
\author{Igor Lesanovsky}
\affiliation{Institut f\"ur Theoretische Physik, Universit\"at T\"ubingen, Auf der Morgenstelle 14, 72076 T\"ubingen, Germany}
\affiliation{School of Physics and Astronomy and Centre for the Mathematics and Theoretical Physics of Quantum Non-Equilibrium Systems, The University of Nottingham, Nottingham, NG7 2RD, United Kingdom}

\begin{abstract}
We introduce and investigate an open many-body quantum system in which kinetically constrained coherent and dissipative processes compete. The form of the incoherent dissipative dynamics is inspired by that of epidemic spreading or cellular-automaton-based computation related to the density-classification problem. It features two non-fluctuating absorbing states as well as a $\mathcal{Z}_2$-symmetric point in parameter space. The coherent evolution is governed by a kinetically constrained $\mathcal{Z}_2$-symmetric many-body Hamiltonian which is related to the quantum XOR-Fredrickson-Andersen model. We show that the quantum coherent dynamics can stabilize a fluctuating state and we characterize the transition between this active phase and the absorbing states. We also identify a rather peculiar behavior at the $\mathcal{Z}_2$-symmetric point. Here the system approaches the absorbing-state manifold with a dynamics that follows a power-law whose exponent continuously varies with the relative strength of the coherent dynamics. Our work shows how the interplay between coherent and dissipative processes as well as symmetry constraints may lead to a highly intricate non-equilibrium evolution and may stabilize phases that are absent in related classical problems.
\end{abstract}

\maketitle

\section{Introduction}

A paradigmatic setting for the study of nonequilibrium phenomena is provided by stochastic processes featuring absorbing states \cite{hinrichsen2000,henkel2008}, i.e., configurations which, once reached by the dynamics, can no longer be left. These systems typically follow elementary rules but display intriguingly complex nonequilibrium behavior. They describe the dynamics of epidemic spreading, the propagation of opinions in a group of voters and also relate to computing tasks, such as the density-classification problem \cite{holley1975,liggett1994,liggett1997,land1995,fuks1997,busic2013}. Despite their microscopic simplicity, systems with absorbing states show phase transitions, even in one dimension, with universal behavior that possesses no counterpart in equilibrium. Already in the classical domain these models are challenging to investigate and analytic solutions remain scarce \cite{hinrichsen2000}. They become even more complex when quantum effects, such as coherence and entanglement, are introduced, which makes them ideal benchmark problems for numerical methods \cite{carollo2019,jo2021} as well as for gauging the capabilities of quantum simulators \cite{zeiher2016,kim2018,browaeys2020,ebadi2021,jo2022}. 

The \emph{directed percolation hypothesis} asserts  that generic (classical) models with a single absorbing state should display emergent physics in the directed percolation universality class \cite{janssen1981,grassberger1982}. This ``rule" is rather general, but it can be broken by introducing additional symmetries. An example is the so-called Domany-Kinzel cellular automaton \cite{domany1984}, which --- at a particularly symmetric point --- features two absorbing states and a universality class known as \emph{compact directed percolation} \cite{hinrichsen2000,henkel2008}. Recently, it has been shown that also quantum effects may alter the universal physics of many-body systems with absorbing states. This was demonstrated in the context of Markovian open quantum systems \cite{griessner2006,diehl2008,kraus2008,diehl2011,tomadin2011,bardyn2013,perez-espigares2017,buca2020} featuring kinetically-constrained dynamics \cite{lesanovsky2013,olmos2014,everest2016,marcuzzi2016,buchhold2017,gutierrez2017,roscher2018,carollo2019,gillman2019,gillman2020,wintermantel2020,helmrich2020,nigmatullin2021,kazemi2021},  which are of interest also in closed-system settings \cite{vanhorssen2015b,lan2018,turner2018,feldmeier2019,pancotti2020}. Recent works have shown that a quantum version of the so-called contact process \cite{hinrichsen2000} --- possessing a single absorbing state --- does not belong in the directed percolation universality class \cite{carollo2019,jo2021} and that quantum effects even allow for a novel type of absorbing-state phase transitions \cite{carollo2022}. 

\begin{figure*}
    \centering
    \includegraphics[width=1\linewidth]{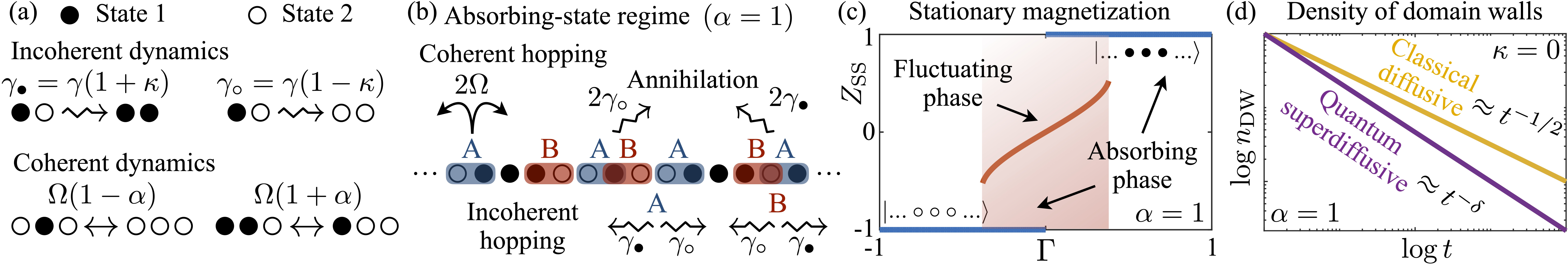}
    \caption{\textbf{Open quantum system with two absorbing states.} (a) Quantum chain made of two-level sites, which can assume the states ${\bullet}$ and ${\circ}$. Each site can undergo a (classical) incoherent state change $\circ/\bullet\rightsquigarrow\bullet/\circ$, occurring a rate $\gamma_{\bullet/\circ}=\gamma(1\pm \kappa)$ per neighbor in the state $\bullet/\circ$ [see Eq.~\eqref{eq:jump_operators}]. Sites can further coherently change their state with rate (``Rabi frequency") $\Omega(1-\alpha)$, if its neighbours are both in the same state, or $\Omega(1+\alpha)$, otherwise [see Eq.~\eqref{eq:Hamiltonian}]. (b) This system is dual to a domain-wall model, with kink ($\circ\bullet$) domain-wall particles A and anti-kink ($\bullet\circ$) domain-wall particles B. For $\alpha=1$ (absorbing-state regime), these can only hop, both coherently and incoherently, and annihilate with rates $2\gamma_{\bullet/\circ}$. (c) For large (in modulus) values of the parameter $\Gamma=\kappa \gamma^2/\Omega^2$ and $\alpha=1$ the stationary state is one of two absorbing states, according to the sign of $\Gamma$. For small $|\Gamma|$, instead, a stationary fluctuating phase emerges, as predicted by Eq.~\eqref{eq:meanfield_eq_state}. (d) The density of domain-wall particles $n_\mathrm{DW}$ [cf.~panel (b)] decays with time $t$ as a power-law for $\alpha=1$ and $\kappa=0$. The decay law varies from a diffusive behavior for $\Omega/\gamma=0$, to a superdiffusive one, $t^{-\delta}$ with $1/2<\delta<1$, for $\Omega/\gamma\neq0$ [see Fig.~\ref{fig:Fig4}(b) below]. }
    \label{fig:Fig1}
\end{figure*}

In this paper, we introduce a kinetically-constrained open quantum system, depicted in Fig.~\ref{fig:Fig1}(a-b), which allows us to investigate how quantum effects impact on the critical behavior of nonequilibrium processes with two absorbing states. By analyzing both stationary and dynamical properties we unveil a rich nonequilibrium phase diagram and obtain two key results. First, we show the existence of a fluctuating phase --- stabilized by quantum effects --- which prevents the system from approaching the absorbing-state manifold [as shown in Fig.~\ref{fig:Fig1}(c)]. Such a novel phase  is not supported by classical dynamics but can solely be observed when quantum coherence exceeds a certain strength. Second, at a $\mathcal{Z}_2$-symmetric point  --- in which the classical dynamics of our model resembles that of the (symmetric) Domany-Kinzel cellular automaton --- we observe critical power-law relaxation towards the absorbing-state manifold. The associated exponent appears to continuously vary from diffusive behavior $t^{-1/2}$ for vanishing quantum coherence --- as expected for compact directed percolation \cite{henkel2008} --- to a \emph{superdiffusive} one \cite{vanhorssen2015} for large coherent rates [as sketched in Fig.~\ref{fig:Fig1}(d)]. Our findings demonstrate that open quantum systems with two absorbing states can display intriguing nonequilibrium physics, where quantum effects lead to novel stationary phases and dynamical behavior. Our numerical results suggest that this phenomenology can be observed even in one-dimensional quantum systems. 

\section{The system}

We consider a quantum chain, with periodic boundary conditions, made of sites that can either be in state $\ket{\bullet}$ or in state $\ket{\circ}$ [cf.~Fig.~\ref{fig:Fig1}(a)]. For convenience, we introduce the Pauli matrices: $\sigma^x=|\bullet\rangle\langle\circ|+|\circ\rangle\langle\bullet|$, $\sigma^y=-i|\bullet\rangle\langle\circ|+i|\circ\rangle\langle\bullet|$ and $\sigma^z=|\bullet\rangle\langle\bullet|-|\circ\rangle\langle\circ|$.

The dynamics of the system state $\rho(t)$ is governed by the quantum master equation \cite{lindblad1976,gorini1976,breuer2002}
\begin{eqnarray}
\dot{\rho}(t)&=&-i[H,\rho(t)]+\mathcal{D}[\rho(t)]. 
\label{eq:master_equation}
\end{eqnarray}
The quantum Hamiltonian is given by  
\begin{equation}
H=\Omega \sum_{k=1}^L K_k \sigma_k^x\, , \, \, \, \mbox{with} \, \,\,  K_k=1-\alpha\,\sigma_{k-1}^z\sigma_{k+1}^z,
\label{eq:Hamiltonian}
\end{equation}
which describes coherent transitions at site $k$ occurring with a rate (``Rabi frequency") which depends on the state of its neighboring sites [cf.~Fig.~\ref{fig:Fig1}(a)], as enforced by the operator $K_k$ for $\alpha>0$. For the special value $\alpha=1$, $K_k$ implements a so-called hard constraint: transitions take place solely when neighbouring sites are in different states, as shown in Fig.~\ref{fig:Fig1}(a), analogously to the so-called XOR-Fredrickson-Andersen model \cite{causer2020}. Note that $H$ possesses a $\mathcal{Z}_2$ symmetry, since it is invariant under the transformation $\sigma_k^x\rightarrow\sigma_k^x$ and $\sigma_k^z\rightarrow -\sigma_k^z$. 

The second contribution in Eq.~\eqref{eq:master_equation} accounts for dissipative classical processes and has the form
\begin{eqnarray}
\mathcal{D}[\rho]&=&\sum_{k,\nu} \left[L_{k,\nu}\rho L^\dagger_{k,\nu}-\frac{1}{2}\left\{L^\dagger_{k,\nu}L_{k,\nu},\rho \right\}\right],
\label{eq:dissipator}
\end{eqnarray}
where the $L_{k,\nu}$ are the so-called jump operators. The rules depicted in Fig.~\ref{fig:Fig1}(a) can be implemented through four types of jump operators ($\nu=\{\circ +, \circ -,\bullet +, \bullet -\}$):
\begin{equation}
L_{k,\bullet \pm}= \sqrt{\gamma_\bullet}\sigma_k^+ n_{k\pm1} \, ,\quad L_{k,\circ \pm}= \sqrt{\gamma_\circ}\sigma_k^- (1-n_{k\pm1}),
\label{eq:jump_operators}
\end{equation}
with $\sigma^\pm=(\sigma^x \pm i\sigma^y)/2$ and $n=\sigma^+\sigma^-=|\bullet\rangle\langle\bullet|$. The jump operators $L_{k,\bullet \pm}$ effectuate the transition $|\circ\rangle\rightsquigarrow|\bullet\rangle$ at site $k$ when the right ($k+1$), respectively, left ($k-1$) neighbor of $k$ is in state $|\bullet\rangle$. Analogously, the operators $L_{k,\circ \pm}$ effectuate the transition $|\bullet\rangle\rightsquigarrow|\circ\rangle$ at site $k$ when the right (left) neighbor of $k$ is in state $|\circ\rangle$ [cf.~Fig.~\ref{fig:Fig1}(a)]. We parametrize the corresponding classical transition rates as $\gamma_{\bullet/\circ}=\gamma(1\pm\kappa)$. 
Here,  $\gamma$ sets the overall rate while $\kappa\, \in\, [-1,1]$ introduces a bias in the different processes, making the transition $|\circ\rangle\rightsquigarrow|\bullet\rangle$ more likely for $\kappa>0$. By construction, the map in Eq.~\eqref{eq:dissipator} possesses the absorbing states $|...\circ\circ\circ...\rangle$ and $|...\bullet\bullet\bullet...\rangle$. For $\alpha=1$ --- in the following referred to as the absorbing-state regime --- these are also eigenstates of the Hamiltonian in Eq.~\eqref{eq:Hamiltonian} so that, in this regime, these are absorbing states for the open quantum dynamics in Eq.~\eqref{eq:master_equation}. We note that the above model may be, in principle, realized with Rydberg atoms, as discussed in Refs.~\cite{kazemi2021,causer2020}.

As sketched in Fig.~\ref{fig:Fig1}(b), the considered quantum system is dual to a domain-wall model \cite{KW1,KW2,kogut1979,ostmann2019}. In the dual lattice formed by bonds (and not by sites), one recognizes two types of particles: particle A representing the (kink) domain wall $\circ \bullet $ and particle B representing the (anti-kink) domain wall $\bullet\circ$. These particles alternate in space, delimiting the extension of clusters of consecutive $\bullet$ or $\circ$ sites. Because of the rules introduced above, such particles can both coherently and incoherently hop, and be annihilated or generated in pairs AB or BA with different rates. Importantly, for $\alpha=1$, their number cannot increase since pairs can only be annihilated and the model becomes a two-species ``reaction-diffusion process" \cite{toussaint1983,hinrichsen2000,tuber2005} with an additional quantum coherent dynamics. In this case, the absorbing states are identified by the states without AB or BA pairs. Thus, the density of domain walls is a valid order parameter in the dual lattice. Details of the mapping are provided in Appendix \ref{App1}. 

\section{Mean-field analysis}

To qualitatively understand the nonequilibrium behavior of our system, we perform a mean-field analysis, expected to be valid in sufficiently high dimensions. The equation of state for the average ``magnetization", $Z=\sum_{k=1}^L \langle \sigma_k^z\rangle/L$, in the mean-field stationary state, which we denote as $Z_\mathrm{SS}$, is [see Appendix \ref{App2} for details]
\begin{eqnarray}
\Omega^2 Z_\mathrm{SS} \left(1-\alpha Z_\mathrm{SS}^2\right)^2&=&\gamma^2\kappa \left(1-Z_\mathrm{SS}^2\right)\, . \label{eq:meanfield_eq_state}
\end{eqnarray}
This equation solely depends on the dimensionless parameter $\Gamma=\kappa\gamma^2/\Omega^2$, quantifying the relative strength between classical and coherent dynamics, modulated by the asymmetry parameter $\kappa$. 

\begin{figure}
    \centering
    \includegraphics[width=1\linewidth]{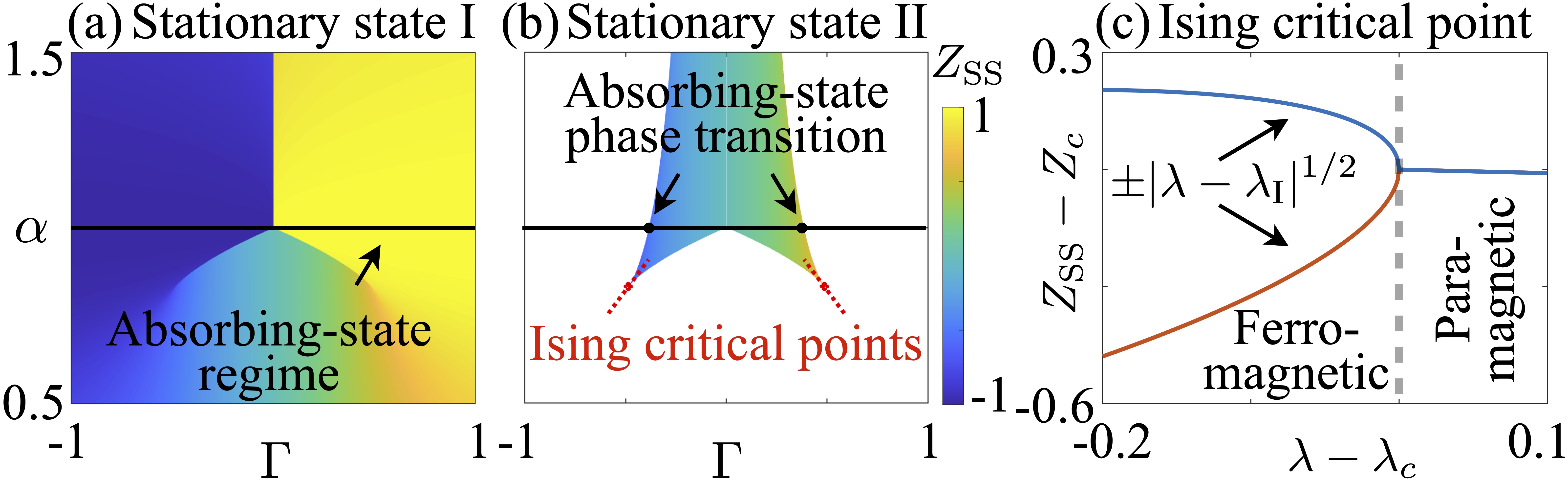}
    \caption{\textbf{Mean-field phase diagram.} (a) Magnetization $Z_\mathrm{SS}$ in stationary state I as a function of $\alpha$ and $\Gamma$. (b) Same as in panel (a) for stationary state II, which also provides the fluctuating phase shown in Fig.~\ref{fig:Fig1}(c) for $\alpha=1$. Highlighted in the plot are the $\alpha=1$ line, the absorbing-state phase transition points [cf.~Fig.~\ref{fig:Fig1}(c)], the two Ising-like critical points, and the associated direction manifesting Ising criticality. (c) Stationary magnetization $Z_\mathrm{SS}$ along the dashed line crossing the Ising critical point at $\Gamma>0$ in panel (b).}
    \label{fig:Fig2}
\end{figure}

For $\alpha=1$, Eq.~\eqref{eq:meanfield_eq_state} features two solutions $Z_\mathrm{SS}=\pm 1$, corresponding to the two absorbing states. However, for $\Gamma<0$ ($\Gamma>0$) only the state with $Z_\mathrm{SS}=-1$ ($Z_\mathrm{SS}=1$) is stable, as shown in Fig.~\ref{fig:Fig1}(c) [cf.~stationary solution I in Fig.~\ref{fig:Fig2}(a)]. For large enough values of $|\Gamma|$, these are the only physical solutions. However, moving within the absorbing-state regime towards $\Gamma\to0$ with $\Omega\neq0$, a stable fluctuating phase emerges at $\Gamma_c=\pm2/(3\sqrt{3})$, as shown in Fig.~\ref{fig:Fig2}(b) and anticipated in Fig.~\ref{fig:Fig1}(c).  The emergence of such ``bi-stable" regime 
is reminiscent of a first-order phase transition, which also appears in the (mean-field) quantum contact process \cite{marcuzzi2016,buchhold2017}.

To get insights into the origin of the nonequilibrium phase transition, we soften the Hamiltonian constraint and expand the analysis beyond the absorbing-state regime, i.e., we consider $\alpha\neq 1$, for which there exist no absorbing states. The phase diagram  in the $\Gamma-\alpha$-plane is shown in Fig.~\ref{fig:Fig2}(a-b). The emergence of the stationary state II, shown in Fig.~\ref{fig:Fig2}(b), is a consequence of the appearance of two critical points, located at 
\begin{eqnarray}
(\pm\Gamma_c,\alpha_c)&=&\left(\pm\frac{4}{5} \sqrt{\frac{8\sqrt{6}-3}{5}},\frac{1}{25}(11+4\sqrt{6})\right), \label{eq:Ising_critical_points}
\end{eqnarray}
with stationary magnetization $Z_c=\pm\sqrt{2\sqrt{2/3}-1}$. To investigate the universal behavior of the system, we perform a perturbative expansion around the critical points. For both points there is a special line in the $\Gamma-\alpha$-plane, parametrized by $\lambda$, along which there appears the breaking of an emergent symmetry \cite{marcuzzi2014}. The order parameter behaves as $|Z_\mathrm{SS}-Z_c|\propto \pm |\lambda_c-\lambda|^{1/2}$, shown in Fig.~\ref{fig:Fig2}(c), with $\lambda_c$ coinciding with the critical point. This is reminiscent of a mean-field Ising model: for $\lambda>\lambda_c$, the system is ``paramagnetic" while for $\lambda<\lambda_c$ two ``ferromagnetic" solutions emerge with a stationary exponent equal to $1/2$. 

\section{ Single-cluster dynamics}

In the following we investigate whether signatures of a  fluctuating phase are already visible in a one-dimensional system. We start by considering the case in which the initial state of the system features a single cluster of sites in $\ket{\bullet}$, or, equivalently, a single pair of A-B domain-wall particles [cf.~Fig.~\ref{fig:Fig1}(b)]. In the absorbing-state regime, the number of clusters cannot increase. The only relevant degree of freedom is thus the cluster length, equal to the number of bonds between particles  A and  B, counted from A onward. 
Such a single-cluster setting reduces to an effective single-body problem, as detailed in Appendix \ref{App3}. 

\begin{figure}[t]
\centering
\includegraphics[width=\linewidth]{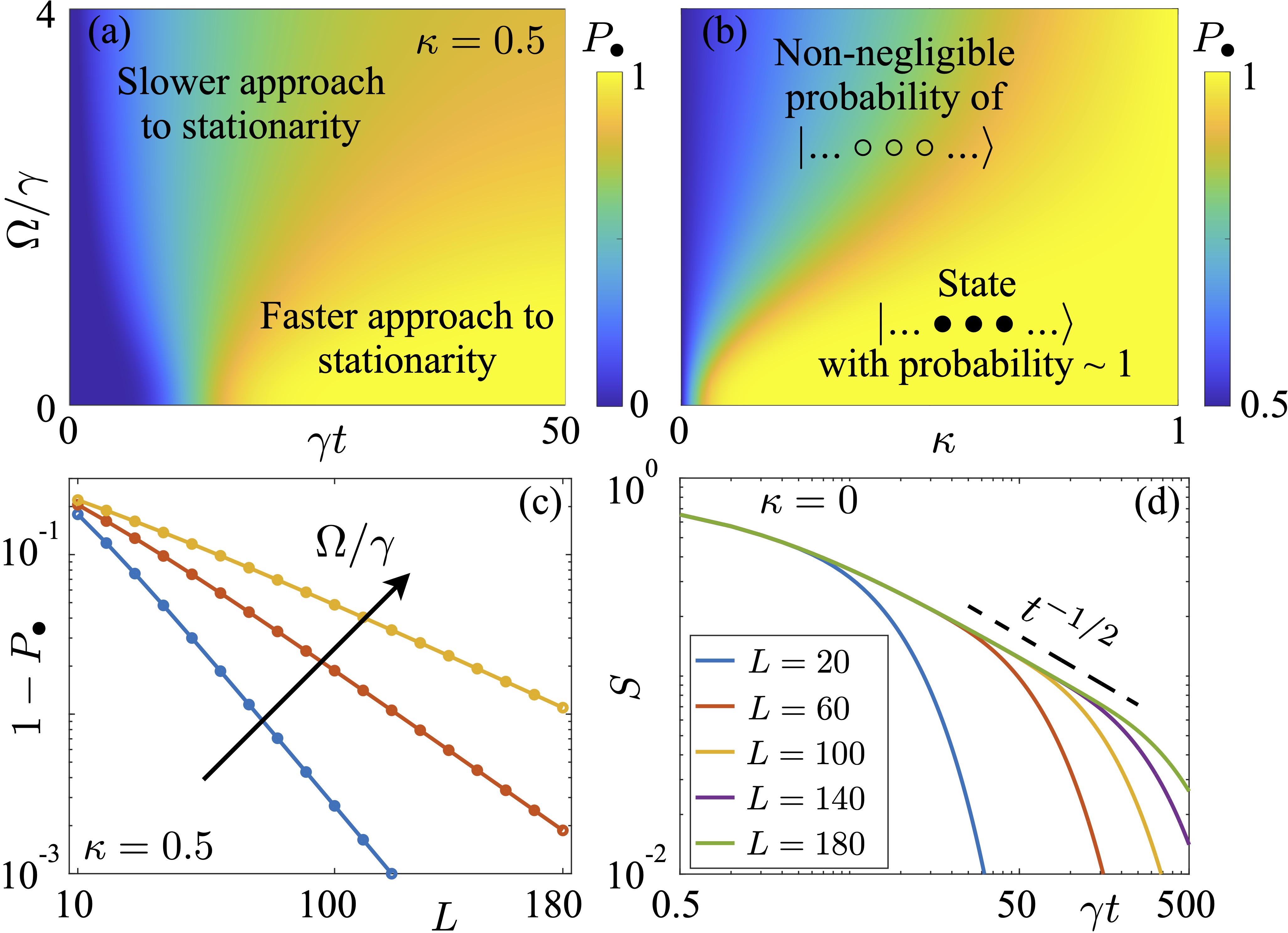}
\caption{{\bf Single-cluster dynamics.} (a) Probability $P_\bullet(t)$ of finding a cluster of length $L$ as a function of time and of $\Omega/\gamma$, for $\kappa=0.5$ and $L=50$. (b) Stationary behavior of $P_\bullet$ for a system of $L=50$ sites, as a function of $\Omega/\gamma$ and $\kappa$. (c) Stationary probability $1-P_\bullet$ as a function of $L$ for $\Omega/\gamma=1.5,2,2.5$ and $\kappa=0.5$. (d) Behavior of the single-seed survival probability $S$ as a function of $t$ for $\Omega/\gamma=1$, $\kappa=0$, and different values of $L$. Upon increasing $L$, a power-law behaviour $\approx t^{-1/2}$ is approached. }
\label{fig:Fig3}
\end{figure}

We consider a system of length $L$ that contains initially a cluster of size $L/2$ ($L$ even), focusing on the regime $\kappa\in[0,1]$. Here we calculate the probability $P_{\bullet}(t)$ for finding that, at time $t$, the cluster has length $L$, i.e., that the system ended up in the absorbing state preferred by the incoherent dynamics for $\kappa>0$ [cf.~Fig.~\ref{fig:Fig1}(a)]. 
As shown in Fig.~\ref{fig:Fig3}(a), the approach to the stationary state is slower the larger the ratio $\Omega/\gamma$ and the stationary value of $P_\bullet$ is smaller, see Fig.~\ref{fig:Fig3}(b). This means that, while the system ends up in the absorbing-state manifold, for increasing $\Omega/\gamma$ the probability of finding the system in state $\ket{...\bullet\bullet\bullet...}$ decreases. This is due to the fact that the coherent dynamics gives rise to symmetric domain-wall hopping. Thus, in the large $\Omega/\gamma$ regime, the asymmetry introduced by $\kappa$ becomes less relevant and there are higher chances that the system approaches the state  $\ket{...\circ\circ\circ...}$. While this may suggest that a fluctuating phase could emerge in the single-cluster setting, the dependence of $P_\bullet$ on the system size $L$ shows that this is not case. The stationary probability $P_\bullet$ indeed converges exponentially to $1$, as shown in Fig.~\ref{fig:Fig3}(c), indicating that even in the presence of quantum effects the system behaves as in the classical limit, $\Omega=0$. 

We now analyze whether quantum effects impact on the dynamics at the $\mathcal{Z}_2$-symmetric point ($\kappa=0$). In the classical limit ($\Omega=0$) the system falls here into the compact directed percolation universality class --- just like the (symmetric) Domany-Kinzel cellular automaton \cite{domany1984}. This can be seen, for instance, by studying the behavior of the \emph{single-seed survival probability}, $S(t)$, defined as the probability of being outside the absorbing-state manifold, at time $t$, when starting from a single site in $\ket{\bullet}$ \cite{hinrichsen2000,henkel2008}. Such a single-seed initial condition has also been widely investigated for the contact process, both in its classical \cite{hinrichsen2000,henkel2008} and, more recently, in its quantum version \cite{gillman2019,jo2021}. For the compact directed percolation universality the probability $S(t)$ displays power-law dynamics $S(t)\approx t^{-1/2}$. As shown in Fig.~\ref{fig:Fig3}(d), this scaling does not appear to change when coherent processes are introduced, i.e., $\Omega$ increases.

\section{ Many-body dynamics in 1D} So far the results suggest that, even in the presence of coherent dynamical processes, the model behaves exactly as in the classical limit, $\Omega=0$. However, we now show that this is not the case and that quantum effects become relevant when considering a genuine many-body setting. We still focus on the absorbing-state regime ($\alpha=1$) and take as initial state the N\'eel state $\ket{\bullet\circ\bullet...\circ}$. To investigate this setting, we use matrix product states and employ a time-evolving-block-decimation algorithm \cite{vidal2003,vidal2004,paeckel2019} that we developed using the package \cite{oseledets2011b} which implements basic algebraic operations \cite{oseledets2011}. We considered system sizes up to $L=14$, with periodic boundary conditions, and used different bond dimensions and discrete time-steps to check consistency of our results.  In particular, we find that a bond dimension $\chi=64$ correctly captures the time evolution of the considered observables for all the cases we have simulated.

\begin{figure}
    \centering
    \includegraphics[width=1\linewidth]{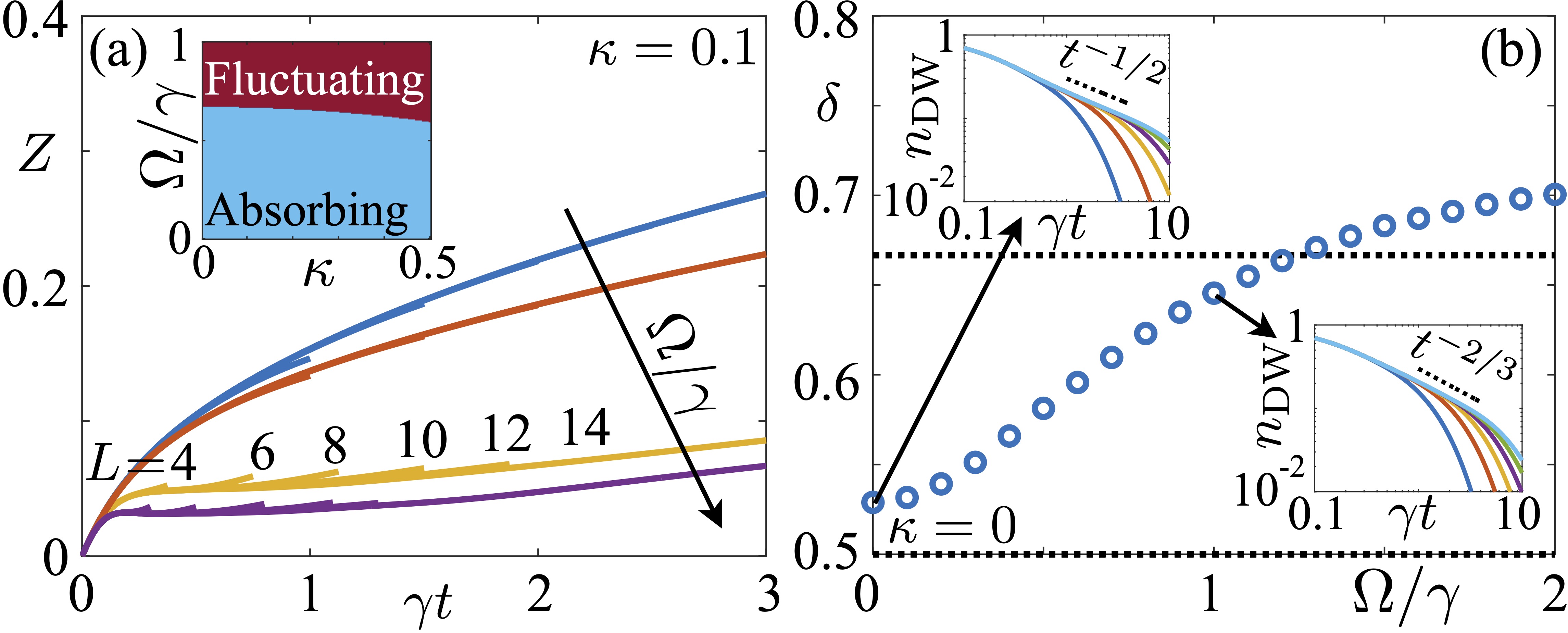}
    \caption{\textbf{Many-body tensor-network simulations.} (a) Magnetization $Z$, starting from the N\'eel state, for $\alpha=1$ and $\kappa=0.1$. The different sets of curves are for $\Omega/\gamma=0,0.5,2,3$. For each $\Omega/\gamma$, we display results for $L=4,6,\dots, 14$. For sufficiently large values of $\Omega/\gamma$, upon increasing the system size $L$, a fluctuating phase in which the magnetization remains small emerges. The inset shows an estimated phase diagram. (b) Estimate --- fit of the numerical data for $\gamma t\in [1, 2]$ --- of the dynamical critical exponent $\delta$  as a function of $\Omega/\gamma$, for $\kappa=0$ and $L=14$. The insets show, in a log-log plot, the domain-wall density,  $n_\mathrm{DW}$, as a function of time, for $L=4,6,\dots 14$, with $\Omega/\gamma=0$ (left) and $\Omega/\gamma=1$ (right). The results  presented are for a bond dimension $\chi=64$. }
    \label{fig:Fig4}
\end{figure}

First, we investigate whether an emergent fluctuating phase [cf.~Fig.~\ref{fig:Fig1}(c)] can be observed in this many-body setting. To this end, we study the time-evolution of the average magnetization $Z$, considering $\kappa>0$. As shown in Fig.~\ref{fig:Fig4}(a), for small values of $\Omega/\gamma$, $Z$ tends to a stationary value associated with an appropriate statistical mixture of the two absorbing states. Upon increasing the system size $L$, the stationary magnetization increases towards the value $1$.  This indicates that the stationary state converges to the absorbing state $\ket{...\bullet\bullet\bullet...}$, which is the one preferred by positive values of $\kappa$. This scenario is identical to that in the classical limit ($\Omega=0$), suggesting that for small values of $\Omega/\gamma$ the Hamiltonian (\ref{eq:Hamiltonian}) acts as a perturbation which merely introduces  an additional diffusive contribution with rate $\Omega^2/\gamma$ \cite{lesanovsky2013} (see also Appendix \ref{App4}). However, when $\Omega/\gamma$ surpasses a ``critical" threshold, a  completely different phenomenology emerges. As shown, e.g., by the curve with $\Omega/\gamma=3$ in Fig.~\ref{fig:Fig4}(a), the system sustains here a ``meta-stable" phase, characterized by small (in modulus) values of the magnetization. In this regime, quantum fluctuations due to the coherent dynamics are strong enough to keep the system away from the absorbing-state manifold. Due to finite-size effects, clearly the system eventually approaches the absorbing-state manifold. However, the trend shown upon increasing $L$, i.e., the emergence of a plateau value for the magnetization and the fact that the meta-stable phase survives for longer times when $L$ is larger,  suggests that such a fluctuating phase may become stable in the limit $L\to\infty$. In the inset of Fig.~\ref{fig:Fig4}(a), we provide an estimated phase diagram showing where the emergence of such a phase may be expected. This is obtained by considering which parameters lead to an initial decrease in the modulus of $Z$ for increasing system sizes.

Finally, we analyze the  $\mathcal{Z}_2$-symmetric point $\kappa=0$ for $\alpha=1$. In this case, the density of domain walls 
\begin{equation}
n_\mathrm{DW}=\frac{1}{L}\sum_{k=1}^L \langle n_{k-1} (1-n_{k})+ (1-n_{k-1})n_{k}\rangle \, ,
\label{eq:density_domain_walls}
\end{equation} 
always decays to zero. Interestingly, however, it does so by showing a power-law decay $n_\mathrm{DW}\approx t^{-\delta}$, which is sustained for longer and longer times for increasing system sizes [cf.~Fig.~\ref{fig:Fig4}(b)]. The power-law exponent $\delta$ continuously varies from the (classical) diffusive value $\delta=1/2$ \cite{hinrichsen2000,henkel2008} to a super-diffusive one ($\delta>1/2$) for increasing $\Omega/\gamma$. The largest $\delta$ is between $1/2$ and $1$ --- also observed in Ref.~\cite{vanhorssen2015} for different models. As shown in Fig.~\ref{fig:Fig4}(b), this value is close to the value $\delta\approx 0.7$, reminiscent of a superdiffusive exponent $2/3$. Quantum effects thus appear to accelerate the domain-wall annihilation process in the $\mathcal{Z}_2$-symmetric regime. 

\section{Discussion} 

We introduced a nonequilibrium quantum system featuring two absorbing states [cf.~Fig.~\ref{fig:Fig1}]. Its mean-field phase diagram in Fig.~\ref{fig:Fig2} displays a bi-stable regime with two possible stationary states. Within the absorbing-state regime, we have shown that sufficiently strong quantum effects can lead to the emergence and the stabilization of a novel fluctuating phase, which is not possible in classical regimes. 
At the $\mathcal{Z}_2$-symmetric point ($\kappa=0$), a fluctuating phase does not exist in one dimension, not even in the presence of quantum effects. On the contrary, the coherent dynamics leads to a speed-up of the (power-law) approach to the absorbing-state manifold. The associated algebraic (power-law) exponent, shown in Fig.~\ref{fig:Fig4}(b), seems to continuously vary with $\Omega/\gamma$. This is reminiscent of the so-called quantum contact process, in which the decay of the particle density at the critical point also follows a power-law with a continuously varying exponent \cite{jo2021}. Moreover, also in that system, the dynamics of a single initial seed (corresponding to our single-cluster dynamics) shows no qualitative deviations from the classical problem, similar to what we observe in our model [cf.~Fig.~\ref{fig:Fig3}(d)].

\acknowledgments
We acknowledge discussions with Juan P.~Garrahan, Gianluca Ceruti, Christian Lubich and Dominik Sulz. The research leading to these results has received funding from the “Wissenschaftler-R\"uckkehrprogramm GSO/CZS” of the Carl-Zeiss-Stiftung and the German Scholars Organization e.V., as well as through the Deutsche Forschungsgemeinsschaft (DFG, German Research Foundation) under Project No. 435696605 and through the Research Unit FOR 5413/1, Grant No. 465199066. G.P. acknowledges support from the Alexander von Humboldt Foundation through a Humboldt research fellowship for postdoctoral researchers.

\appendix

\section{Mapping to the domain wall model}
\label{App1}
In this section, we explain the mapping of the system discussed in the main text, see Eqs.~\eqref{eq:master_equation}-\eqref{eq:jump_operators}, to a model of domain walls with the Kramiers-Wannier duality transformation, see, e.g., Ref.~\cite{KW1,KW2,kogut1979,ostmann2019}. In particular, for the sake of simplicity in the explanation of the mapping, we follow the notation convention of Ref.~\cite{ostmann2019}, where a chain of $L+2$ sites with open boundary conditions is considered. The two spins at the boundary points of the chain $0$ and $L+1$ are fixed to be down. 

Within the duality transformation a one-dimensional ``dual lattice'' is associated to the original one-dimensional spatial lattice. Sites of the dual lattice correspond to the bonds of the original lattice (and vice versa). On each site $k$ of the dual lattice a set of dual spin operators $\mu^{\alpha}_{k}$ ($\alpha=x,y,z$) is defined as 
\begin{equation}
\begin{split}
&\sigma^x_{k}=\mu_k^x\mu_{k+1}^x, \quad \sigma^y_k=(-1)^{k+1}\left(\prod_{l=1}^{k-1}\mu_l^z\right)\mu_k^y\mu_{k+1}^x, \\
&\sigma^z_k=(-1)^{k+1}\prod_{l=1}^k\mu_l^z,
\end{split}
\label{eq:KW_duality_1}
\end{equation}
with the inverse transformation given by 
\begin{equation}
\begin{split}
&\mu_k^x=\left(\prod_{l=1}^{k-1} \sigma^x_l  \right), \quad \mu^y_k=-\sigma^y_{k-1}\sigma^z_k \left(\prod_{l=1}^{k-2}\sigma^x_l  \right) ,\\
&\mu_k^z=-\sigma^z_k\sigma^z_{k-1}.    
\end{split}
\label{eq:KW_duality_2}
\end{equation}
Note that Eqs.~\eqref{eq:KW_duality_1} and \eqref{eq:KW_duality_2} apply to $L+2$ spin operators $\sigma^{\alpha}_k$ defined on the original lattice sites $k=0\dots L+1$ with two fictitious down spins at the boundary points of the chain $0$ and $L+1$, as anticipated at the beginning of this section.
One has therefore $L+1$ dual lattice operators $\mu^{\alpha}_k$ defined on the dual lattice sites (the bonds of the original lattice) $k=1\dots L+1$. At the boundary dual lattice sites, one accordingly has $\mu_1^{z}=\sigma_1^z$ and $\mu_{L+1}^z=\sigma_L^z$. From Eqs.~\eqref{eq:KW_duality_1} and \eqref{eq:KW_duality_2}, it is also simple to check that the operators $\mu^{\alpha}_{k}$ can be considered as Pauli spin operators since they satisfy the same Pauli spin algebra as the corresponding $\sigma^{\alpha}_{k}$ operators. We note that $\mu^z_k$ in Eq.~\eqref{eq:KW_duality_2} feels whether neighbouring spins are aligned or not and therefore whether a domain wall is present on the bond $k$ or not. The total number of domain walls operator $N_{\mathrm{DW}}$ is accordingly given by
\begin{equation}
 N_{\mathrm{DW}}=\sum_{k=2}^{L} \frac{1+\mu_{k}^z}{2}.
\label{eq:DW_number}
\end{equation}
Note that Eq.~\eqref{eq:DW_number} involves a sum over the $L-1$ bulk sites of the dual lattice (thereby excluding the two fictitious spins at the boundary sites $0$ and $L+1$) as a consequence of the open boundary conditions adopted in this section. From the third equation in \eqref{eq:KW_duality_2}, it is, however, immediate to verify that the density of domain walls $n_{\mathrm{DW}}=(L-1)^{-1} N_{\mathrm{DW}}=(L-1)^{-1}\sum_{k}(1+\mu_k^z)/2$ coincides with the expression in Eq.~\eqref{eq:density_domain_walls} of the main text (up to a boundary term, irrelevant in the thermodynamic limit $L\to \infty$, coming from the choice of periodic boundary conditions taken in Eq.~\eqref{eq:density_domain_walls}). 
It is, moreover, important to emphasize that $\mu_k^z$ is invariant under $\mathcal{Z}_2$ transformations and, as a matter of fact, it does not distinguish a kink ($\circ\bullet$ -- A particle -- see Fig.~\ref{fig:Fig1} in the main text) from an anti-kink ($\bullet\circ$, B particle). 
The operator $\mu^x_k$, instead, flips all the spins to the left of the lattice site $k$, and, therefore, it creates a domain wall upon acting on a state with all the spins pointing upwards (or downwards). 

The mapping of the model in Eqs.~\eqref{eq:master_equation}-\eqref{eq:jump_operators} in terms of the dual spin operators in Eqs.~\eqref{eq:KW_duality_1} and \eqref{eq:KW_duality_2} is useful as it sheds light on the emergent physics of the open quantum system in terms of domain wall particles hopping (both coherently and incoherently) and pairwise annihilating. We start by writing the Hamiltonian in Eq.~\eqref{eq:Hamiltonian} with the dual spin operators as
\begin{equation}
\begin{split}
H&=\Omega \sum_{k=1}^{L} \mu_k^x\mu_{k+1}^x +\alpha \mu^y_{k}\mu^y_{k+1}\\
&=\sum_{k=1}^{L} J\left(\frac{1+\zeta}{2}\right)\mu_k^x\mu_{k+1}^x +J\left(\frac{1-\zeta}{2}\right) \mu^y_{k}\mu^y_{k+1},
\end{split}
\label{eq:Hamiltonian_dual_spin}
\end{equation}  
where $J\geq 0$ parameterizes the ferromagnetic interaction constant and $\zeta$ the anistropy between the couplings in the $x$ and the $y$ directions. In particular $J$ and $\zeta$ are function of $\Omega$ and $\alpha$ as    
\begin{equation}
J=\Omega(1+\alpha), \quad \zeta=\frac{1-\alpha}{1+\alpha}.    
\label{eq:xy_couplings}    
\end{equation}
The Hamiltonian in Eqs.~\eqref{eq:Hamiltonian_dual_spin} and \eqref{eq:xy_couplings} is readily recognized as the $XY$ spin chain with open boundary conditions, which can be mapped to a free fermionic theory, see, e.g., Ref.~\cite{lieb1961two}. In the case $\alpha=1$, where the constraint in the Hamiltonian \eqref{eq:Hamiltonian} becomes hard, $\zeta=0$ and equation \eqref{eq:Hamiltonian_dual_spin} reduces to the $XX$ spin chain, as remarked also in Ref.~\cite{ostmann2019}. The transverse magnetization $M=\sum_k \mu_k^z$ of the $XY$ chain is equal (up to a constant) to the total number of domain walls $N_{\mathrm{DW}}$ in Eq.~\eqref{eq:DW_number}. The latter is, consequently, conserved only for $\alpha=1$, i.e., in the case of the $XX$ spin chain Hamiltonian. In the case $\alpha \neq 1$ of the $XY$ model, as a matter of fact, domain walls can be created or annihilated in pairs and therefore only their parity is conserved under the Hamiltonian time evolution. The states with zero number of domain walls, $\ket{... \circ\circ\circ ...}$ and $\ket{...\bullet\bullet\bullet ...}$, are therefore stationary states of the Hamiltonian only for $\alpha=1$, as stated in the main text.

To proceed with the mapping of the open quantum system considered in the main text to the domain wall picture we need to consider the dissipative part of the dynamics in Eqs.~\eqref{eq:dissipator} and \eqref{eq:jump_operators}. The four types of jump operators in Eq.~\eqref{eq:jump_operators} residing at each lattice $k$ in the domain wall picture are written as 
\begin{widetext}
\begin{subequations}
\begin{align}
L_{k,\bullet,+}&=\sqrt{\gamma_\bullet}\sigma_k^+ n_{k+1}  =\sqrt{\gamma_{\bullet}} \left(\frac{1+S_{k+1}}{2}\right)(\mu^{-}_k\mu^{-}_{k+1}+\mu^{+}_k \mu^{-}_{k+1}), 
\label{eq:jump_operators_DW_1} \\
L_{k,\bullet,-}&=\sqrt{\gamma_\bullet}\sigma_k^+ n_{k-1}=\sqrt{\gamma_{\bullet}} \left(\frac{1+S_{k-1}}{2}\right)(\mu^{-}_k\mu^{-}_{k+1}+\mu^{-}_{k}\mu^{+}_{k+1}), 
\label{eq:jump_operators_DW_2} \\
L_{k,\circ,+}&=\sqrt{\gamma_\circ}\sigma_k^- (1-n_{k+1})=\sqrt{\gamma_{\circ}} \left(\frac{1-S_{k+1}}{2}\right)(\mu^{-}_k\mu^{-}_{k+1}+\mu^{+}_k \mu^{-}_{k+1}), 
\label{eq:jump_operators_DW_3} \\
L_{k,\circ,-}&=\sqrt{\gamma_\circ}\sigma_k^- (1-n_{k-1})=\sqrt{\gamma_\circ}\left(\frac{1-S_{k-1}}{2}\right)(\mu^{-}_k\mu^{-}_{k+1}+\mu^{-}_{k} \mu^{+}_{k+1}), 
\label{eq:jump_operators_DW_4}
\end{align}
\label{eq:jump_operator_all_dual}%
\end{subequations}
\end{widetext}
with $\mu^{\pm}_k=(\mu_k^x\pm i\mu^y_k)/2$ the raising-lowering operator for the dual lattice spin operators. The rates are parametrized as in the main text $\gamma_{\bullet/\circ}=\gamma(1\pm\kappa)$. 
In Eq.~\eqref{eq:jump_operator_all_dual}, $S_{k}$ denotes a string operator
\begin{equation}
S_k= (-1)^{k+1}\prod_{l=1}^k\mu^z_l=\sigma^z_k, \quad \mbox{with} \quad S_k^2=1. 
\label{eq:string_operator}
\end{equation}
One realizes that each of the jump operators describes a superposition of a AB (or BA) pair destruction ($\mu^{-}_k\mu^{-}_{k+1}$) process and incoherent hopping of particles A-kink (or B, anti-kink) to the right ($\mu^{-}_{k} \mu^{+}_{k+1}$) or to the left ($\mu^{+}_k \mu^{-}_{k+1}$), as shown in Fig.~\ref{fig:Fig1}(b) of the main text. Crucially, these processes happen with different rates ($\gamma_{\bullet}$ and $\gamma_{\circ}$) depending on the pair annihilated (AB or BA) and on the particle hopping being A or B and on the direction of the hopping. The dual operators $\mu^{\pm}_k$, as stated before, do not, however, distinguish particles A from the B ones. This is accomplished by the factors in Eq.~\eqref{eq:jump_operator_all_dual} containing the string operator $S_k$ \eqref{eq:string_operator}, which constrain the hopping depending on the magnetization of the left or right neighbouring spin and therefore it identifies the nature of the particle involved in the process being A or B. Note that the operator $S_k$ is non-local in terms of the dual lattice operators $\mu^z_k$, since the latter is not sensitive to the direction of the magnetization, but only to the relative alignment of neighboring spins. 

The jump operators in Eq.~\eqref{eq:jump_operator_all_dual} cause the number of domain walls to decrease in time. The vacuum states of domain walls $\ket{... \bullet\bullet\bullet ...}$ and $\ket{... \circ\circ\circ ...}$ are, however, stationary states of the full master equation \eqref{eq:master_equation} only for $\alpha=1$. For $\alpha \neq 1$, as a matter of fact, the Hamiltonian can create pairs of A and B particles and the stationary state can, consequently, exhibit a finite density of domain walls.

\section{From the Heisenberg equations to the mean-field equation of state for the magnetization}
\label{App2}
In this section, we show how to obtain, within a mean-field analysis the equation of state for the stationary magnetization presented and discussed in the main text. 

The starting point is the calculation of the Heisenberg equations of motion for the single-site spin operators. The equations are the following 
\begin{widetext}
\begin{subequations}
\begin{align}
\partial_t \sigma_m^x&=2\alpha\Omega\sigma_m^y\left(\sigma_{m-2}^z\sigma_{m-1}^x+\sigma_{m+1}^x\sigma_{m+2}^z\right)-\left(\gamma_\bullet+\gamma_\circ\right)\sigma_m^x,  \, \\
\partial_t \sigma_m^y&= -2\Omega \sigma_m^z - 2\alpha\Omega\left[\sigma_{m}^x\left(\sigma_{m-2}^z\sigma_{m-1}^x+\sigma_{m+1}^x\sigma_{m+2}^z\right)-\sigma_{m-1}^z\sigma_{m}^z\sigma_{m+1}^z\right]-\left(\gamma_\bullet+\gamma_\circ\right)\sigma_m^y, \, \\
\partial_t \sigma_m^z&=2\Omega \sigma_m^y\left[1-{\alpha}\sigma_{m-1}^z\sigma_{m+1}^z\right]-\left(\gamma_\bullet+\gamma_\circ\right)\left[\sigma_m^z-\frac{1}{2}\left(\sigma_{m-1}^z+\sigma_{m+1}^z\right)\right]+\left(\gamma_\bullet-\gamma_\circ\right)\left[1-\frac{\sigma_m^z}{2}\left(\sigma_{m-1}^z+\sigma_{m+1}^z\right)\right]\, . \label{eq_SM_Heiseberg_3}
\end{align}
\label{eq_SM:Heisenberg}%
\end{subequations}
\end{widetext}
In passing, we note from Eq.~\eqref{eq_SM_Heiseberg_3} that the magnetization $\sum_{k=1}^L \sigma^z_k/L$ is conserved at the $\mathcal{Z}_2$ symmetric point ($\kappa=0$) only in the case the Rabi frequency $\Omega$ is set to zero. Then, we take the expectation value of the above equations and perform a so-called mean-field decoupling of the correlation functions, e.g., $\langle \sigma_m^\alpha \sigma_k^\beta\rangle \approx \langle \sigma_m^\alpha\rangle \langle\sigma_k^\beta\rangle$. Further assuming a homogeneous initial state, which amounts to $\langle \sigma_k^\alpha\rangle =\langle \sigma_h^\alpha\rangle$, $\forall k,h$, we can introduce the variables $X\equiv\langle \sigma_k^x\rangle$, $Y\equiv\langle \sigma_k^y\rangle$ and $Z\equiv\langle \sigma_k^z\rangle$. Within such a homogeneous mean-field approximation, the following set of dynamical mean-field equations are obtained 
\begin{subequations}
\begin{align}
\dot{X} &= 4\alpha\Omega\,X\,Y\, Z - 2 \gamma X \, ,\\
\dot{Y}  &= -2\Omega Z - 2\alpha \Omega Z\left(2X^2-Z^2\right) - 2\gamma Y\, ,\\
\dot{Z} &= 2\Omega Y\left(1-\alpha Z^2\right) + 2\gamma \kappa \left(1-Z^2\right).
\end{align}
\end{subequations}
To obtain the stationary state of the system, we need to set the left-hand sides of the above equations to zero. This yields $X_\mathrm{SS}=0$, $Y_\mathrm{SS}=-\frac{\Omega}{\gamma} Z_\mathrm{SS} \left(1-\alpha Z_\mathrm{SS}^2\right)$ as well as the equation of state for the magnetization $Z_\mathrm{SS}$
\begin{eqnarray}
\Omega^2 Z_\mathrm{SS} \left(1-\alpha Z_\mathrm{SS}^2\right)^2&=&\gamma^2\kappa \left(1-Z_\mathrm{SS}^2\right)\, , \label{eq_SM:meanfield_eq_state}
\end{eqnarray}
which is Eq.~\eqref{eq:meanfield_eq_state} of the main text.

\section{Effective single-cluster model} 
\label{App3}
As mentioned in the main text, the single-cluster setting can be investigated through an effective single-body dynamical model. 
This can be obtained by defining the states $\ket{m}$,
with $m=0,1,\dots L$ denoting the extension of the cluster. 

According to the dynamics of the domain-wall particles sketched in Fig.~\ref{fig:Fig1}(b), the open quantum dynamics for the extension of the cluster can be constructed as follows. For the coherent dynamics we take the Hamiltonian 
\begin{equation}
H=4\Omega \sum_{m=1}^{L-2} \big(\ket{m}\!\bra{m+1}+\ket{m+1}\!\bra{m}\big)\, .
\label{Ham-simple-model}
\end{equation} 
The above dynamics encodes the coherent hopping of the domain walls, which results in a change of the size of the clusters. The factor $4\Omega$ comes from the fact that the cluster can coherently increase (decrease) because of a kink A particle moving to the left (right), with rate $2\Omega$ or because of an anti-kink particle B moving towards the right (left), also with rate $2\Omega$. The combination of these two effects, which can be seen also in the sketched in Fig.~\ref{fig:Fig1}(b), gives a rate $4\Omega$. The extremal values of the sum appearing in Eq.~\eqref{Ham-simple-model} reflect the presence of  the absorbing states $H\ket{0}=H\ket{L}=0$. The incoherent dynamics is instead accounted for by a dissipative Lindblad contribution  characterised by two sets of jump operators.  The first consists of jump operators that can only decrease the number of occupied sites. These are  
\begin{equation}
J_m^\circ=\sqrt{2\gamma_{\circ}} \ket{m}\!\bra{m+1}\, ,
\label{jump_down}
\end{equation}
for $m=0,1,\dots L-2$. Here the factor $2\gamma_\circ$ is also due to the fact that an incoherent decrease of  the size of the cluster can be achieved both with the kink particle A jumping to the right and with an anti-kink particle B jumping to the left. The second set consists instead of jump operators which increase the number of occupied sites in the central region. These are, in analogy to the previous ones, given by the following jump operators
\begin{equation}
J_m^\bullet=\sqrt{2\gamma_\bullet} \ket{m+1}\!\bra{m}\, ,
\label{jump_up}
\end{equation}
for $m=1,2,\dots L-1$. Note that also jump operators have been chosen in such a way that state $\ket{0}$ and state $\ket{L}$ are absorbing states for the effective model dynamics. 

The overall dynamics of the single cluster is thus implemented by the following quantum master equation evolving the density matrix of the system, $\rho_t$, as
\begin{equation}
\begin{split}
\dot{\rho}_t=&-i[H,\rho_t]+\sum_{m=1}^{L-1} \left(J_m^\bullet\rho_t J_m^{\bullet \dagger}-\frac{1}{2}\left\{J_m^{\bullet\dagger} J_m^\bullet,\rho_t\right\}\right)\\
&+\sum_{m=0}^{L-2} \left(J_m^\circ\rho_t J_m^{\circ \dagger}-\frac{1}{2}\left\{J_m^{\circ\dagger} J_m^\circ,\rho_t\right\}\right)\, .
\end{split}
\label{QME}
\end{equation}

\section{Perturbative contribution of the quantum coherent dynamics}
\label{App4}
In this section, we show how the quantum dynamics provides an additional dissipative contribution in the limit $\Omega\ll\gamma$. Here, the Hamiltonian (\ref{eq:Hamiltonian}) is considered as perturbation to the generator of the dissipative dynamics \eqref{eq:dissipator}. 

To derive the effective classical dynamics, we introduce the projector 
\begin{eqnarray}
\mathcal{P}[\rho]=\sum_\mathcal{C}|\mathcal{C}\rangle\langle\mathcal{C}|\rho|\mathcal{C}\rangle\langle\mathcal{C}|,
\end{eqnarray}
which projects a given density matrix $\rho$ onto the classical basis formed by configurations $|\mathcal{C}\rangle$. Using second order perturbation theory \cite{lesanovsky2013}, the density matrix projected onto the classical subspace, $\mu(t)=\mathcal{P}[\rho(t)]$, evolves according to
\begin{eqnarray}
\!\!\!\!\!\!\!\partial_t \mu(t)=\mathcal{D}[\mu(t)]\!+\!\int_0^\infty \!\!\!\!dt'\, \mathcal{P} \mathcal{H}\mathcal{Q}\exp\left(\mathcal{D}t' \right)\mathcal{Q}\mathcal{H} [\mu(t)],
\end{eqnarray}
where $\mathcal{Q}=1-\mathcal{P}$ is the complement of $\mathcal{P}$ and we have defined $\mathcal{H}[O]=-i[H,O]$. Such an equation can be obtained by employing standard Nakajima-Zwanzig projector techniques \cite{breuer2002}, keeping all terms up to second-order in the coherent rate $\Omega$ and performing a Markovian approximation. The latter is needed to obtain a time-independent generator. Evaluating  the second term yields
\begin{eqnarray}
\!\!\!\!\!\!\!\partial_t \mu(t)=\mathcal{D}[\mu(t)]+\frac{\Omega^2}{\gamma}\sum_k K_k^2\left[\sigma_k^x \mu(t) \sigma_k^x - \mu(t) \right]. \label{eq:classical_limit}
\end{eqnarray}
This shows that in the perturbative limit the Hamiltonian introduces single spin-flips that take place at the constrained rate $(\Omega^2/\gamma)K^2_k$.
Equation \eqref{eq:classical_limit} can be solved efficiently using classical Monte-Carlo methods. However, here we simply want to note that, in the regime $\alpha=1$, the coherent hopping of domain-wall particles implemented by the Hamiltonian is effectively acting as an incoherent one in the limit $\Omega\ll\gamma$. Moreover, since the rate of the Hamiltonian transition is independent on whether the final state is $\bullet$ or $\circ$, the resulting incoherent hopping of domain-wall particles is symmetric. As it happens for stochastic exclusion processes, a symmetric incoherent hopping of these particles is expected to give rise to diffusion of domain walls. This is in contrast with the dissipative channels described by $\mathcal{D}$: when $\kappa\neq0$ these have a preferred direction and thus give rise to an overall ballistic motion of domain walls [cf.~Fig.~\ref{fig:Fig1}(b)]. 

\bibliography{biblio}

\end{document}